# An Equality Set Projection Approach for TSO-DSO Coordination Dispatch

Bo Li, *Member, IEEE*, Xicong Pang, *Student Member, IEEE*, Guangrui Wei, Haiwang Zhong, *Senior Member, IEEE*, *Grant Ruan, Member, IEEE,* Zhengmao Li, *Member, IEEE*, Edris Pouresmaeil, *Senior Member, IEEE*

***Abstract*—Coordinated optimization dispatch (COD) of transmission system operator (TSO) and distribution system operator (DSO) can effectively ensure system security and efficiency under high-penetration distributed energy resource (DER) integration. Researches of large-scale COD problem can be categorized into iterative approaches that allow DSO to dispatch independently, and non-iterative methods based on projections of feasible regions (FR). However, the iterative methods suffer from low computational convergence and efficiency, while non-iterative methods struggle to solve equivalent projections with high-dimensional FR. To address these issues, this paper proposes a TSO-DSO coordinated dispatch approach based on an accelerated non-iterative Equality Set Projection (ESP) algorithm. First, ESP algorithm is employed to overcome the bottleneck of high-dimensional FR construction. Second, an $\ell 2$ regularization-based accelerated method is proposed to reduce computational burden when degeneracy occurs. Accelerated ESP algorithm constructs projection of FR via adjacent facet searching. Therefore, it is less sensitive to the increase of vertices and could efficiently construct the projection of high-dimensional FR. Case studies on a polyhedron dataset, IEEE 33-Bus System and T118D10 TSO-DSO system demonstrate the effectiveness and computational efficiency of the proposed COD approach.**

NOMENCLATURE

| | |
|---|---|
| *Indices and sets* | |
| T/D | Superscript markers for distinguishing parameters/variables of the TSO and DSO |
| $\Omega^{\mathrm{Th,T}}$, $\Omega^{\mathrm{RE,T}}$, $\Omega^{\mathrm{LN,T}}$ | Sets of thermal generators, renewable generators, and transmission lines in TSO |
| $\Omega^{\mathrm{Th,D}}$, $\Omega^{\mathrm{RE,D}}$, $\Omega^{\mathrm{LN,D}}$ | Sets of thermal generators, renewable generators and transmission lines in DSO |
| $\Omega^{\mathrm{T}} / \Omega^{\mathrm{D}}$ | Set of buses in TSO and DSO |
| $\Omega_i^{\mathrm{G,T}} / \Omega_i^{\mathrm{G,D}}$ | Set of generators connected to bus *i* in TSO and DSO |
| $\Omega_i^{\mathrm{T}} / \Omega_j^{\mathrm{T}}$ | Sets of transmission lines whose “sending” node are bus *i* and *j* |
| $\Omega_i^{\mathrm{D}} / \Omega_j^{\mathrm{D}}$ | Sets of distribution lines whose “sending” node are bus *i* and *j* |
| $\Omega^{\mathrm{Time}}$ | Sets of discrete-time horizon |
| $g$ | Index of generator |
| $i, j$ | Index of bus |
| $k$ | Index of segments used in the piecewise-linear approximation |
| $t$ | Index of discrete-time horizon |
| *Parameters* | |
| $c_g^{\mathrm{T}}(\cdot) / c_g^{\mathrm{D}}(\cdot)$ | Generation cost function of thermal generators in TSO/DSO |
| $c^{\mathrm{RE}}$ | Unit penalty cost for curtailment renewable generation |
| $c^{\mathrm{LS}}$ | Unit penalty cost for load shedding |
| $c^{\mathrm{Ex}}$ | Unit power exchange cost at interface between TSO and DSO |
| $\underline{P}_{i,j}^{\mathrm{L,T}} / \overline{P}_{i,j}^{\mathrm{L,T}}$ | Lower and upper bounds of the power transfer on line (*i, j*) |
| $\underline{P}_i^{\mathrm{Tie}} / \overline{P}_i^{\mathrm{Tie}}$ | Lower and upper bounds of the power transfer at the TSO-DSO interconnection |
| $\underline{P}_g^{\mathrm{Th,T}} / \overline{P}_g^{\mathrm{Th,T}}$ | Upper and lower bound of thermal generator output in TSO |
| $\overline{r}_i^{\mathrm{LS,T}} / \overline{r}_i^{\mathrm{LS,D}}$ | Load shedding limit of bus *i* in TSO and DSO |
| $\overline{P}_g^{\mathrm{RE,T}}$ | Maximum forecast output of renewable generator |
| $P_g^{\mathrm{RAMP,T}}$ / $P_g^{\mathrm{RAMP,D}}$ | Ramp rate of thermal unit in TSO and DSO |
| $P_{i,t}^{\mathrm{T}}$ | Active power demand at bus *i in TSO* |
| $P_{i,t}^{\mathrm{D}} / Q_{i,t}^{\mathrm{D}}$ | Active / reactive demands at bus *i in DSO* |
| $\overline{P}_g^{\mathrm{Th,D}}$ | Upper bound of thermal generator *g* in DSO |
| $g_{i,j} / b_{i,j}$ | Conductance and susceptance of line (*i*, *j*) |
| $\underline{\theta}_i^{\mathrm{T}} / \overline{\theta}_i^{\mathrm{T}}$ | Lower and upper bounds of the bus voltage angle in DSO |
| $\underline{v}_i^{\mathrm{T}} / \overline{v}_i^{\mathrm{T}}$ | Lower and upper bounds of the squared voltage magnitude at bus *i* |
| $\underline{\theta}_i^{\mathrm{D}} / \overline{\theta}_i^{\mathrm{D}}$ | Lower and upper bounds of the bus voltage angle in DSO |
| $\underline{v}_i^{\mathrm{D}} / \overline{v}_i^{\mathrm{D}}$ | Lower and upper bounds of the squared voltage magnitude at bus *i* |
| $S_{i,j,t}^{\mathrm{D}}$ | Maximum apparent power of line (*i*, *j*) |
| $K$ | The number of segments used in the piecewise-linear approximation |
| *Variables* | |
| $r_{g,t}^{\mathrm{RE,T}} / r_{g,t}^{\mathrm{RE,D}}$ | Renewable curtailment of generator *g* at time *t* in TSO and DSO |
| $r_{i,t}^{\mathrm{LS,T}} / r_{i,t}^{\mathrm{LS,D}}$ | Load shedding at bus *i* at time *t* in TSO and DSO |

This work was supported in part by the Guangxi Natural Science Foundation (2025GXNSFBA069263). (Corresponding author: Haiwang Zhong.)
Bo Li, Xicong Pang and Guangrui Wei are with the School of Electrical Engineering, Guangxi University, Nanning 530004, China (e-mail: boli@gxu.edu.cn; pangxicong@st.gxu.edu.cn, 1152376170@qq.com).
Bo Li, Haiwang Zhong are with the Department of Electrical Engineering, Tsinghua University, Beijing 100084, China (e-mail: zhonghw@tsinghua.edu.cn).
Edris Pouresmaeil and Zhengmao Li are with the Department of Electrical Engineering and Automation, Aalto University, Espoo 02150, Finland(e-mail: edris.pouresmaeil@aalto.fi, lizh0049@e.ntu.edu.sg;
Grant Ruan is with the Laboratory of Information and Decision Systems, Massachusetts Institute of Technology, Cambridge, MA 02139 USA (e-mail: gruan@mit.edu)

| | |
|---|---|
| $P_{g,t}^{\mathrm{T}}$ | Active power output of generator $g$ at time $t$ in TSO |
| $P_{g,t}^{\mathrm{D}} / Q_{g,t}^{\mathrm{D}}$ | Active/reactive power output of generator $g$ at time $t$ in DSO |
| $P_{i,j,t}^{\mathrm{L,T}}$ | Power flow between bus $i$ and $j$ |
| $P_{i,t}^{\mathrm{Tie}}$ | Power exchange at the interface between TSO and DSO |
| $P_{i,j,t}^{\mathrm{L,D}} / Q_{i,j,t}^{\mathrm{L,D}}$ | Active/reactive power flows on line $(i, j)$ at time $t$ |
| $S_{i,j,t}^{\mathrm{L,D}}$ | Apparent power flow on line $(i, j)$ at time $t$ |
| $\theta_{i,t}^{\mathrm{T}} / \theta_{j,t}^{\mathrm{T}}$ | Voltage angles of bus $i$ and $j$ at time t |
| $\theta_{i,t}^{\mathrm{D}} / \theta_{j,t}^{\mathrm{D}}$ | Voltage angles of bus $i$ and $j$ at time t |
| $V_{i,t} / V_{j,t}$ | Bus voltage magnitudes |

# I. INTRODUCTION

The structure of modern power system is constructed hierarchically by multi-voltage-level distribution system and transmission system. With the increasing number of DERs and the integration of multiple energy sectors, systems in different regions and levels are coupled tighter. Meanwhile, the large number of DERs enables flexible dispatch by the DSO. Therefore, COD between different subsystems is growing in prominence to guarantee the reliability, robustness and economy of the entire power system. For example, Federal Energy Regulatory Commission (FERC) Order 2222 underscores that transmission–distribution coordination in dispatch encompasses not only the management of active and reactive power flows at TSO-DSO interfaces, but also the integration of distributed energy resources' flexibility, visibility, and controllability. However, as separate entities, the TSO and DSO need to ensure the privacy and security of their internal data, which become an obstacle to COD. Additionally, TSO-DSO coordination in large-scale power system could lead to high computational burden [1].

Two types of methods have been proposed to solve this problem: 1) iterative method and 2) non-iterative method. Iterative method decouples the COD problem into several sub-problems with reduced dimensionality and communication burden. These problems will be solved independently, and an overall optimum solution will be achieved via the iterative information exchange between the different hierarchies. In detail, representative iterative methods include the alternating direction method of multipliers [2], Benders decomposition [3], [4], Lagrange relaxation [5], [6] and cutting plane method [7]. Ref.[8] proposes a unified scheme to quantify time-coupled flexibilities at the distribution level for TSO–DSO coordination and develops a DSO-centric optimization with value/price allocation that internalizes intertemporal DER behaviors. Ref.[9] proposes a coordinated frequency-constrained stochastic economic dispatch model for integrated transmission–distribution systems and develops a two-layer distributed ADMM-based framework to handle joint chance constraints. Ref.[5] proposes an operational-planning TSO–DSO coordination model and develops an ℓ1-proximal surrogate Lagrange relaxation to efficiently decompose and solve coupled Optimal Power Flow/Unit Commitment-type constraints across transmission and distribution. Ref.[3] proposes a distributed approach utilizing logic-based Benders decomposition for the COD problem in an integrated electric and heating system. Ref.[4] proposes a decentralized multi-area dynamic economic dispatch framework based on modified generalized Benders decomposition to achieve fast convergence without parameter tuning under a hierarchical coordinator architecture. Despite the iterative COD methods demonstrate certain effectiveness in alleviating the computational burden, these methods necessitate extensive information exchange and face computational challenges. On the one hand, these methods inevitably encounter obstacles such as poor convergence and poor scalability. On the other hand, frequent exchange of information between different hierarchies will cause heavy communication burdens.

Due to the inherent limitations of iterative COD methods, many non-iterative approaches have been developed. One type of methods develops models with fixed shape FR to reduce computational burden caused by time-coupling constraints. For example, hypercubes [10] and fixed convex polyhedral [11] are employed to represent the model parameter and could be aggregated through Minkowski sum. Ref.[12], [13], [14] propose a non-iterative method based on the equivalent projection(EP) to solve the large-scale COD problem. This method projects the reduced FR of the DSO onto the optimal dispatch model of the TSO, thereby reducing the complexity of TSO-DSO coordinated dispatch without revealing private information. Based on this principle, an EP based framework is developed in Ref.[12], [13]to capture technical characteristic of DSO. The efficacy of these methods is contingent upon the efficient acquisition of the FR. Existing methods for constructing the FR includes network equivalence methods [15], [16], Fourier-Motzkin elimination (FME) [17], and optimization based methods represented by vertex enumerations(VE) [12], [13], [14], [18]. Network equivalence methods, such as Ward equivalence [16] and Thevenin [15], are widely adopted in the early stages of TSO-DSO coordination research. Nonetheless, these methods encounter a bottleneck in optimal dispatch since operation limits, and the cost function are not considered. Fourier-Motzkin elimination iteratively reduces dimensionality of the model by eliminating internal variable in the constraints. However, this method leads to a large amount of redundant constraints [19] that leads to high computational complexity. The above methods are difficult to be applied to high-dimensional and multi-period models, because curse of dimensionality will occur when the number of variables increases.

Table I. REVIEW OF COORDINATED OPTIMAL DISPATCH METHODS

| Method | Non-iteration | Decomposed | Multi-level coordination | Communication burden | Computational burden |
|---|---|---|---|---|---|
| Benders decomposition [3] | × | ✓ | × | High | Medium |
| Lagrange relaxation [5],[6] | × | ✓ | × | Medium | Medium |
| Cutting plan method [7] | × | ✓ | × | High | Medium |
| Ward Equivalence [15] | ✓ | × | × | Low | Medium |
| Thevenin Equivalence [16] | ✓ | × | × | Low | Medium |
| FME+ADMM [17] | ✓ | ✓ | ✓ | Low | High |
| Equivalent Projection Framework | | | | | |
| PVE [12],[13] | ✓ | ✓ | ✓ | Low | Medium (high-dimension) |
| Quickhull-PVE [14] | ✓ | ✓ | ✓ | Low | Medium (high-dimension) |
| FAPVE [18] | ✓ | ✓ | ✓ | Low | Medium (high-dimension) |
| **Accelerated ESP (This paper)** | ✓ | ✓ | ✓ | Low | Low (high-dimension) |

The progressive vertex enumerations (PVE) algorithm is developed in Ref.[12], [13] to efficiently construct the FR of DSO. The basic idea of the PVE algorithm is to identify the vertices that are critical to the overall shape of FR. Then, new vertices can be found by expanding the convex hull of existing vertices. Ref.[14] enhances the PVE algorithm to effectively construct high-dimensional flexible regions of DSO by adopting Quickhull algorithm. Further, Ref.[18] refines the Quickhull algorithm and proposes a fast approximation PVE (FAPVE) algorithm. However, as the dimension of the FR increases, the number of vertices could grow exponentially, which increases the computational burden of PVE algorithm.

This paper proposes a non-iterative COD approach. An accelerated ESP algorithm is developed to construct the FR projection of high-dimensional DSO model via adjacent facet searching process. Further, an $\ell 2$ regularization-based accelerated method is proposed and applied to the ESP algorithm to enhance its computational efficiency under dual degenerate conditions. The contributions are as follows:

i) A scalable TSO-DSO coordination approach is proposed to enhance computational efficiency in solving TSO-DSO coordination problems. The reduced FR of DSO is projected into the model of TSO, enabling TSO to solve TSO-DSO coordination problems independently without iteration.

ii) An ESP Algorithm is proposed for the EP calculation in power system applications. To eliminate the computational bottleneck causes by cure of dimensionality, ESP Algorithm construct FR by facet-to-facet searching, bypassing the exponential explosion of vertex enumeration.

iii) An accelerated method is developed to enhance the computational efficiency of ESP algorithm. An $\ell 2$ regularization term is added to the objective function to ensure its convexity, thereby preventing dual degeneracy and the associated recursive computations.

The rest of this paper is organized as follows. Section II introduces the TSO-DSO coordination model. Section III presents the accelerated ESP algorithm. Section IV is case study. Section V is conclusion.

## II. TSO-DSO Coordination Model

In this section, the mathematical formulation of the DSO problem and the TSO problem are presented in Section II.A and Section II.B, respectively. The overall non-iterative TSO-DSO coordination approach is developed in Section II.C.

### A. TSO Model

Suppose a TSO with interconnected several DSOs via substations, the ideal TSO-DSO coordination is that a single entity can fully model, observe, and dispatch resources and networks in TSO and DSO systems. However, it is unrealistic within the wholesale market. The decentralized approach of TSO-DSO is more practical, in which the TSO is responsible for economic dispatch of transmission energy resources (TERs) and aggregated DER. The DSO is responsible for validating, aggregating, and dispatching DER service bids.

*Objective*: TSO aims at minimizing total generation costs of all TERs, load shedding costs, interface power exchange costs, and renewable curtailment penalties.

$$C_{\mathrm{T}} = \min \left( \begin{array}{l} \sum_{g \in \Omega^{\mathrm{Th,T}}} c_g^{\mathrm{T}} \left( P_{g,t}^{T} \right) + \sum_{i \in \Omega^{\mathrm{LN,T}}} c^{\mathrm{Ex}} P_{i,t}^{\mathrm{Tie}} + \\ \sum_{i \in \Omega^{\mathrm{T}}} c^{\mathrm{LS}} r_{i,t}^{\mathrm{LS,T}} + \sum_{g \in \Omega^{\mathrm{RE,T}}} c^{\mathrm{RE}} r_{g,t}^{\mathrm{RE,T}} \end{array} \right) \tag{1}$$

The TSO operation should meet the following constraints:

$$\begin{array}{l} \sum_{g \in \Omega_i^{\mathrm{G,T}}} P_{g,t}^{\mathrm{T}} - \sum_{g \in \Omega_i^{\mathrm{RE,T}}} r_{g,t}^{\mathrm{RE,T}} = \sum_{(i,j) \in \Omega_i^{\mathrm{T}}} P_{i,j,t}^{\mathrm{L,T}} - \sum_{(i,j) \in \Omega_j^{\mathrm{T}}} P_{i,j,t}^{\mathrm{L,T}} + P_{i,t}^{\mathrm{T}} \\ + P_{i,t}^{\mathrm{Tie}} - r_{i,t}^{\mathrm{LS,T}} \;,\;\; i \in \Omega^{\mathrm{T}},\; t \in \Omega^{\mathrm{Time}} \end{array} \tag{2}$$

$$\left| P_{i,j,t}^{\mathrm{L,T}} \right| \le b_{ij} \left( \theta_{i,t} - \theta_{i,t} \right), \forall (i,j) \in \Omega^{\mathrm{LN,T}}, t \in \Omega^{\mathrm{Time}} \tag{3}$$

$$\underline{P}_{i,j}^{\mathrm{L,T}} \le P_{i,j,t}^{\mathrm{L,T}} \le \overline{P}_{i,j}^{\mathrm{L,T}} \tag{4}$$

$$\underline{P}_{i}^{\mathrm{Tie}} \le P_{i,t}^{\mathrm{Tie}} \le \overline{P}_{i}^{\mathrm{Tie}} \tag{5}$$

$$\underline{P}_{g}^{\mathrm{Th,T}} \le P_{g,t}^{\mathrm{T}} \le \overline{P}_{g}^{\mathrm{Th,T}} \;,\;\; g \in \Omega^{\mathrm{Th,T}} \tag{6}$$

$$-P_{g}^{\mathrm{RAMP,T}} \le P_{g,t}^{\mathrm{T}} - P_{g,t-1}^{\mathrm{T}} \le P_{g}^{\mathrm{RAMP,T}}, \;\; g \in \Omega^{\mathrm{Th,T}} \tag{7}$$

$$0 \le r_{g,t}^{\mathrm{RE,T}} \le P_{g,t}, g \in \Omega^{\mathrm{RE,T}}, t \in \Omega^{\mathrm{Time}} \tag{8}$$

$$0 \le r_{i,t}^{\mathrm{LS,T}} \le \overline{r}_{i}^{\mathrm{LS,T}}, i \in \Omega^{\mathrm{T}}, t \in \Omega^{\mathrm{Time}} \tag{9}$$

$$\underline{\theta}_{i}^{\mathrm{T}} \le \theta_{i,t}^{\mathrm{T}} \le \overline{\theta}_{i}^{\mathrm{T}}, \quad i \in \Omega^{\mathrm{T}},\; t \in \Omega^{\mathrm{Time}} \tag{10}$$

where constraint (2) represents the active power balance at each bus considering interface exchange power. Constraint (3) represents DC line flow relation. The capacity limitation for tie-line is limited by constraint (5). Constraint (6) represents power generation limit of generation units. Constraint (7) represents ramping-rate limits of thermal generation units. Constraint (8) represents generation curtailment limits of renewable energy generators. Constraint (9) represents load shedding limits. Constraint (10) represents phase angle limitations at each bus.

### B. DSO Model

Suppose a DSO with many DERs and connected to TSO via transformer, DSO is responsible for validating DER service bids and then managing DERs to ensure system operational security.

*Objective*: DSO aims to minimize the total generation costs, load shedding costs, DREs curtailments penalties, and interface power exchange costs.

$$C_{\mathrm{D}} = \min\left( \begin{array}{l} \sum_{g\in\Omega^{\mathrm{Th,D}}} c_g^{\mathrm{D}}\left(P_{g,t}^{\mathrm{D}}\right) + \sum_{i\in\Omega^{\mathrm{LN,D}}} c^{\mathrm{Ex}} P_{i,t}^{\mathrm{Tie}} + \\ \sum_{i\in\Omega^{\mathrm{D}}} c^{\mathrm{LS}} r_{i,t}^{\mathrm{LS,D}} + \sum_{g\in\Omega^{\mathrm{RE,D}}} c^{\mathrm{RE}} r_{g,t}^{\mathrm{RE,D}} \end{array} \right) \tag{11}$$

Assuming DSO has a radial topology, based on linearized AC power flow model of distribution network systems. The linearized DistFlow model is employed in DSO model, and thus the following approximation is made for bus voltages:

$$v_{i,t} = V_{i,t}^2,\ v_{j,t} = V_{j,t}^2 \tag{12}$$

The linearized operation constraints of DSO are presented as follows:

$$\begin{array}{l} \sum_{g\in\Omega_i^{\mathrm{G,D}}} P_{g,t}^{\mathrm{D}} + P_{i,t}^{\mathrm{Tie}} - \sum_{g\in\Omega_i^{\mathrm{RE,D}}} r_{g,t}^{\mathrm{RE,D}} = \sum_{(i,j)\in\Omega_i^{\mathrm{D}}} P_{i,j,t}^{\mathrm{L,D}} \\ - \sum_{(i,j)\in\Omega_j^{\mathrm{D}}} P_{i,j,t}^{\mathrm{L,D}} + P_{i,t}^{\mathrm{D}} - r_{i,t}^{\mathrm{LS,D}},\ \ i\in\Omega^{\mathrm{D}},\ t\in\Omega^{\mathrm{Time}} \end{array} \tag{13}$$

$$\begin{array}{l} \sum_{g\in\Omega_i^{\mathrm{G,D}}} Q_{g,t}^{\mathrm{D}} = \sum_{(i,j)\in\Omega_i^{\mathrm{D}}} Q_{i,j,t}^{\mathrm{L,D}} - \sum_{(i,j)\in\Omega_j^{\mathrm{D}}} Q_{i,j,t}^{\mathrm{L,D}} \\ \qquad + Q_{i,t}^{\mathrm{D}},\qquad i\in\Omega^{\mathrm{D}},\ t\in\Omega^{\mathrm{Time}} \end{array} \tag{14}$$

$$P_{i,j,t}^{\mathrm{L,D}} = g_{i,j}\frac{v_{i,t}-v_{j,t}}{2} - b_{i,j}(\theta_{i,t}^{\mathrm{D}} - \theta_{j,t}^{\mathrm{D}}), i,j\in\Omega^{\mathrm{D}},\ t\in\Omega^{\mathrm{Time}} \tag{15}$$

$$Q_{i,j,t}^{\mathrm{L,D}} = -b_{i,j}\frac{v_{i,t}-v_{j,t}}{2} - g_{i,j}(\theta_{i,t}^{\mathrm{D}} - \theta_{j,t}^{\mathrm{D}}), i,j\in\Omega^{\mathrm{D}},\ t\in\Omega^{\mathrm{Time}} \tag{16}$$

$$\begin{array}{l} P_{i,j,t}^{\mathrm{L,D}}\cos\frac{2k\pi}{K} + Q_{i,j,t}^{\mathrm{L,D}}\sin\frac{2k\pi}{K} \le S_{i,j,t}^{\mathrm{L,D}}\cos\frac{\pi}{K}, \\ i,j\in\Omega^{\mathrm{D}}, t\in\Omega^{\mathrm{Time}}, k=1,2,...K \end{array} \tag{17}$$

$$(\underline{v}_i^{\mathrm{D}})^2 \le v_{i,t} \le (\overline{v}_i^{\mathrm{D}})^2\ ,\quad i\in\Omega^{\mathrm{D}},\ t\in\Omega^{\mathrm{Time}} \tag{18}$$

$$\underline{\theta}_i^{\mathrm{D}} \le \theta_{i,t}^{\mathrm{D}} \le \overline{\theta}_i^{\mathrm{D}},\quad i\in\Omega^{\mathrm{D}},\ t\in\Omega^{\mathrm{Time}} \tag{19}$$

$$0 \le P_{g,t}^{\mathrm{D}} \le \overline{P}_g^{\mathrm{Th,D}},\quad g\in\Omega^{\mathrm{Th.D}}\ , t\in\Omega^{\mathrm{Time}} \tag{20}$$

$$-P_g^{\mathrm{RAMP,D}} \le P_{g,t}^{\mathrm{D}} - P_{g,t-1}^{\mathrm{D}} \le P_g^{\mathrm{RAMP,D}},\ g\in\Omega^{\mathrm{Th,D}} \tag{21}$$

$$0 \le r_{g,t}^{\mathrm{RE,D}} \le P_{g,t}^{\mathrm{D}}, g\in\Omega^{\mathrm{RE,D}}, t\in\Omega^{\mathrm{Time}} \tag{22}$$

$$0 \le r_{i,t}^{\mathrm{LS,D}} \le \overline{r}_{i,t}^{\mathrm{LS,D}}, i\in\Omega^{\mathrm{D}}, t\in\Omega^{\mathrm{Time}} \tag{23}$$

where the constraints (13)-(14) are active and reactive power balance constraints of the DSO, respectively. Based on the linearized DistFlow model and second-order Taylor expansion, linearized power flow of lines in the DSO are formulated as constraints (15)–(16). The line power limitation is expressed via a piecewise linearized form in constraint (17) [20]. Constraints (18)-(19) represent the limits of voltage amplitude and phase angle for each bus, respectively. Constraints (20)–(21) represent the power output limitation and ramping-rate limitation of thermal generators. Renewable energy curtailment limits and load shedding limits are represented by constraints (22)-(23), respectively.

### C. TSO-DSO Coordination Approach

As each of the TSO-DSO coordination models differ in coordination way, hierarchical optimization is more suitable, as shown in Fig. 1. In this way, a large-scale optimization problem is decomposed into many subproblems, which are organized in a hierarchical structure. Each subproblem optimizes local decisions and submits information to the next hierarchical level. The details are summarized as follows:

Step 1: DER sends bids to the DSO.

Step 2: The DSO performs the economic dispatch to seek optimal solution of DER bids while meeting operating security constraints. The aggregated bids are submitted to TSO.

Step 3: The TSO performs the economic dispatch to seek optimal solutions and sends commands to the DSO.

Step 4: The DSO send commands to DERs to follow interface power exchange requirement.

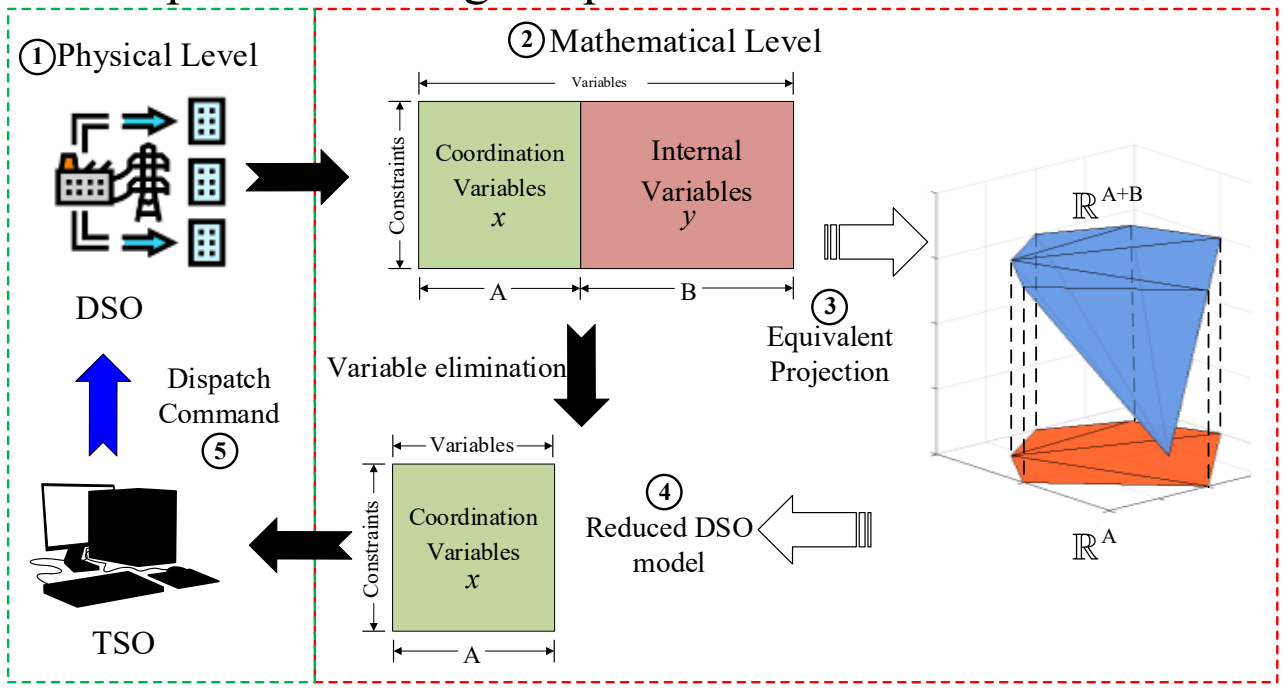

Fig.1. TSO-DSO coordination structure

The TSO-DSO coordination problem is formulated combined TSO model and DSO model as shown in follows:

$$\min_{P,r,X} C_{\mathrm{T}} + C_{\mathrm{D}} \tag{24}$$

$$s.t., \text{(2)-(10),(13)-(23)} \tag{25}$$

where interface power exchange is limited by tie-lines in constraint (5).

It is noted that the non-linear AC power flow and voltage constraints of TSO and DSO could be solved via introduction of linearized power flow approach, which has good performance in computation efficiency and accuracy.

This paper focuses on how to efficiently solve TSO-DSO coordination model via EP approach, detailed in Section III and IV. The proposed EP approach can transform a high-dimensional original FR of DSO to a low-dimensional FR, which is more convenient for submitted to the TSO. Then, the TSO could perform economic dispatch solution and send them to the DSO, achieving a non-iterative solution for TSO-DSO coordination.

## III. The Accelerated ESP Algorithm for TSO-DSO Coordination

### A. DSO Equivalent Projection

Suppose a DSO with many DERs connected with the TSO via feeders or tie-lines, originally flexible region of DSO is a high-dimensional convex hull due to temporal coupling constraints of devices, network transmission constraint, and power balance constraint. According to the Minkowski-Weyl Theorem, convex polyhedron can be represented by the convex combination of a finite number of vertices and the intersection of a finite number of hyperplane [21]. The basic idea of vertex based approach is to search for all vertex according to PVE algorithm, converting vertex representation to hyperplane representation [12], [13]. TSO prefers to obtain low-dimensional boundary information as external parameters of TSO model, but it fails to provide cost information of DSO, which is necessary for its economic performance. Consequently, this paper extends boundary information by considering internal cost information of DSO to make it more practical for DSO dispatch.

The feasible region of DSO is formulated as a convex polyhedron:

$$\Phi_1 = \{(x,y) \in \mathbb{R}^d \times \mathbb{R}^k \mid Ax + By \le d\} \tag{26}$$

where $x$ denotes the coordination variable including power at the tie-line between TSO and DSO, and power outputs of controllable devices. $y$ represents internal variables including voltage magnitudes, phase angles, and active/reactive power flows. $A$, $B$, and $d$ are the coefficient matrixes.

According to the definition of equivalent projection, the high-dimensional FR of DSO can be projected to a low-dimensional FR, formulated as follow:

$$\Phi_2 = \{(x \in \mathbb{R}^d \mid Gx \le b\} \tag{27}$$

where $G$ and $b$ are the coefficient matrixes.

It is noted that convex polyhedron $\Phi_1$ and $\Phi_2$ are linear, this transformation can be understood as the projection of convex polyhedron onto a subspace, which is capable of reflecting its internal technical and economic operation characteristics [22].

The operational cost of DSO depends on outputs of controllable devices. However, the number of coordination variables in $\Phi_2$ will influence computational efficiency of EP. Thus, this paper leverages the Epigraph reformulation to converts the internal terms of objective function (11) into to a an upper bound:

$$\sum_{i\in\Omega^{\mathrm{D}}} c^{\mathrm{LS}} r_{i,t}^{\mathrm{LS,D}} + \sum_{g\in\Omega^{\mathrm{RE,D}}} c_i^{\mathrm{RE}} r_{i,t}^{\mathrm{RE,D}} + \sum_{g\in\Omega^{\mathrm{Th,D}}} c_g^{\mathrm{D}}\left(P_{g,t}^{\mathrm{D}}\right) \le \pi_e \tag{28}$$

where $\pi_e$ is a continuous variable. In this setting, the objective function (11) is reformulated as follow:

$$C_{\mathrm{D}} = \sum_{i\in\Omega^{\mathrm{LN,D}}} c^{\mathrm{Ex}} P_{i,t}^{\mathrm{Tie}} + \pi_e \tag{29}$$

The equivalence between objective function (11) and (29) is ensured by inequality (28). Therefore, the dimensionality of the projected FR of DSO is reduced by simplifying internal terms of DSO model.

### *B. Equality Set Projection Algorithm*

This subsection provides overview of ESP algorithm. Firstly, the notion of an equality set and the projection properties will be introduced. Then, the ESP algorithm will be demonstrated.

**Lemma 1**. A convex polyhedron $P$ can be expressed as the interaction of a finite number of closed halfspace:

$$P \triangleq \{z \in \mathbb{R}^n \mid Az \le b\}, A \in \mathbb{R}^{q\times n}, b \in \mathbb{R}^q \tag{30}$$

Further, a facet $F$ which belongs to $P$ could be express as the interaction of $P$ and an affine hull $\{z \in \mathbb{R}^n \mid a^T z = b\}$:

$$F = P \cap \{z \in \mathbb{R}^n \mid a^T z = b\}, a \in \mathbb{R}^n, b \in \mathbb{R} \tag{31}$$

**Lemma 2**. Any two facets (($n$−1)-faces) of a polyhedron are either not joined in the face or are connected by the inclusion of exactly one edge. Two faces $F_1$ and $F_2$ of $P$ are regarded as adjacent if the intersection $F_1 \cap F_2$ is a facet of both.

**Definition 1**. If $P$ is defined by the interaction of $q$ halfspaces and $E \subseteq \{1,\ldots,q\}$, the set E is an equality set of $P$ if and only if $\mathrm{E} = G(E)$.

$$G(E) \triangleq \{i \in \{1,\ldots,q\} \mid A_i z \le b_i, \forall z \in P_{\mathrm{E}}\} \tag{32}$$

$$P_{\mathrm{E}} \triangleq P \cap \{z \mid A_{\mathrm{E}} z \le b_{\mathrm{E}}\} \tag{33}$$

where the rows of $A_{\mathrm{E}}$ are the rows of $A$ whose indices are in E. The rows of $b_{\mathrm{E}}$ are the rows of $b$ whose indices are in E.

**Fact 1.** If the set E is an equality set of the polyhedron $P$, then $\mathrm{aff}\, P_{\mathrm{E}} = \{z \in \mathbb{R}^n \mid A_{\mathrm{E}} z \le b_{\mathrm{E}}\}$ and $\dim P_{\mathrm{E}} = n - \mathrm{rank} \mathrm{A}_{\mathrm{E}}$.

**Lemma 3.** If $E$ is an equality set of the polyhedron $P$, then $P_{\mathrm{E}}$ is a face of $P$. If $F$ is a face of $P$, then there exists a unique equality set $E$ such that $F = P_{\mathrm{E}}$.

**Definition 2**. If $P \subset \mathbb{R}^d \times \mathbb{R}^k$ is a polyhedron, then the projection of $P$ onto $\mathbb{R}^d$ is expressed:

$$\begin{aligned}\pi_d(P) &\triangleq \{x \in \mathbb{R}^d \mid \exists y \in \mathbb{R}^k, (x,y) \in P\} \\ &= \{x \in \mathbb{R}^d \mid Gx \le g\}\end{aligned} \tag{34}$$

$$P \triangleq \{(x,y) \in \mathbb{R}^d \times \mathbb{R}^k \mid Cx + Dy \le b\} \tag{35}$$

where $C \in \mathbb{R}^{q\times d}$, $D \in \mathbb{R}^{q\times k}$, $b \in \mathbb{R}^q$.

The goal is to compute the matrix $G$ and vector $g$, such that the projection $\pi_d(P)$ can be obtained.

**Fact 2.** If $F_1,\ldots,F_i$ are the facets of a polyhedron $P$, and $F_i = \{x \in \mathbb{R}^n \mid a_i^T x \le b_i\}$, then

$$P = \mathrm{aff}\, P \cap \left\{ x \in \mathbb{R}^n \left| \begin{bmatrix} a_1^T \\ \vdots \\ a_i^T \end{bmatrix} x \le \begin{bmatrix} b_1 \\ \vdots \\ b_i \end{bmatrix} \right. \right\} \tag{36}$$

**Lemma 4.** If $M \triangleq \{(x,y) \in \mathbb{R}^d \times \mathbb{R}^k \mid C_{\mathrm{E}} x + D_{\mathrm{E}} y \le b_{\mathrm{E}}\}$ is an affine set, then the projection of $M$ onto $\mathbb{R}^d$ can be written as:

$$\pi_d(M) \triangleq \left\{ x \in \mathbb{R}^d \left| \mathrm{N}\left(D_{\mathrm{E}}^T\right)^T C_{\mathrm{E}} x = \mathrm{N}\left(D_{\mathrm{E}}^T\right)^T b_{\mathrm{E}} \right. \right\} \tag{37}$$

where the columns of matrix $\mathrm{N}(A)$ form an orthonormal basis for $\mathrm{null}(A)$.

Overall, the matrix $G$ and the vector $g$ are formed from the affine hulls of the facets of the projection. Each facet of $F$ of the projection $\pi_d(P)$ is defined by a unique equality set E of $P$. The ESP algorithm searches for unique equality sets.

The ESP algorithm is initialized by finding a random facet $F$ of $\pi_d(P)$ (*Initialization*). The equality sets that define facets of $P$ that project to the ridges of $\pi_d(P)$ that are subsets of $F$ can then be computed (*Ridge Search*). A list consists of every ridge discovered. Recall Lemma 2, each ridge is contained in exactly two facets of $P$. In each iteration, given selected ridge-facet list element, the equality set of the second containing facet can be computed (*Adjacent Facet Search*).

Recall **Lemma 2**, each ridge is visited exactly twice. Thus, by removing every ridge that appears in the list twice at the end of each iteration, it is guaranteed that the algorithm has a finite execution time. In the non-degenerate case, the complexity of ESP algorithm is linear in the number of facets [23]. The iterative step will stop when the list is empty, in which all facets will have been found.

Once all the equality sets that define facets of P that project to facets of $\pi_d(P)$ have been collected, the matrix $G$ and the vector $g$ can be computed by **Fact 2** and **Lemma 4**.

The ESP algorithm process consists of three main steps, as shown in **Algorithm 1**.

1) *Initialization*: The input is polyhedron *P*. The output is random equality set $E_0$ of $P$, then $\pi_d\left(P_{E_0}\right)$ is a facet of the projection $\pi_d\left(P\right)$. We chose random direction $\gamma \in \mathbb{R}^d$ and solve the linear programming (LP) (38) by moving alone the ray $\gamma r$ until the point $(\gamma r^*, y^*)$ on the projected boundary is located.

$$\begin{aligned} &\max_{r,y}\ r \\ &s.t.\ C\gamma r + Dy \le b \\ &(r,y) \in \mathbb{R}\times\mathbb{R}^k,\ r \ge 0 \end{aligned} \tag{38}$$

where $\gamma \in \mathbb{R}^d$ is a random vector. $r$ is a positive value. The equality set $E_0 = \left\{i \middle| C_i\gamma r^* + D_i y^* = b_i\right\}$ of $P$ associated with point $(\gamma r^*, y^*)$ is identified. $\left[a_f^T \ \ b_f\right]$ can be computed by **Lemma 4**:

$$\left[a_f^T \ \ b_f\right] \leftarrow \mathrm{N}\left({D_{E_0}}^T\right)^T \left[C_{E_0} \ \ b_{E_0}\right] \tag{39}$$

Thus, a facet $(E_0, a_f, b_f)$ of $\pi_d\left(P\right)$ is obtained.

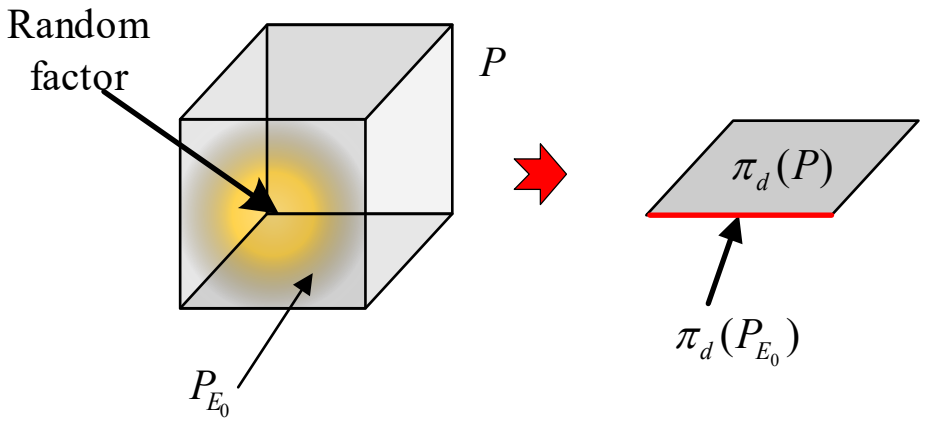

Fig.2. Process of Initialization

2) *Ridge Search (RS)*: The input is an equality set that defines a facet $\pi_d\left(P_E\right) = \left\{x \middle| a_f^T x = b_f\right\} \cap \left\{x \middle| Sx \le t, S \in \mathbb{R}^{p\times d}, t \in \mathbb{R}^p\right\}$ of the projection $\pi_d\left(P\right)$. The output is the equality sets of the ridges of $\pi_d\left(P_E\right)$, which can be written as $\pi_d\left(P_E\right) \cap \left\{x \middle| S_i x = t_i, i = 1,\ldots,p\right\}$. This step shows how to compute $S$ and $t$.

The dimension of the face $P_E$ is $d-1$. Thus, all ridges of $\pi_d\left(P_E\right)$ can be written as:

$$\pi_d\left(P_{E\cup\{i\}}\right) = \left\{x \middle| a_f^T x = b_f, S_i x = t_i, Sx \le t\right\}, i \in E^c \tag{40}$$

where $E^c$ is hyperplane sets except for $E_0$.

For each facet *F* of $\pi_d\left(P_E\right)$, there exists an $i \in E^c$ such that $F = \pi_d\left(P_{E\cup\{i\}}\right)$. Therefore, $\pi_d\left(P_{E\cup\{i\}}\right)$ is a facet of $\pi_d\left(P_E\right)$ only if $\dim \pi_d\left(P_{E\cup\{i\}}\right) = \dim \pi_d\left(P_E\right) - 1 = d-2$. Equivalently, the rank is two.

$$Q\left(i\right) \triangleq \left\{ j \in E^c \middle| \operatorname{rank}\begin{bmatrix} a_f^T & b_f \\ S_{\{i,j\}} & t_{\{i,j\}} \end{bmatrix} = 2 \right\} \tag{41}$$

The equality set of the ridge $\pi_d\left(P_{E\cup\{i\}}\right)$ is given by the set $Q\left(i\right)$, in which the LP (42) is proposed to test if $Q\left(i\right)$ is satisfied in (41). If there exists an $x$ and a strictly negative $\tau$ that satisfies LP (42), then (41) is satisfied. Therefore, $\left[a_r^T \ b_r\right] \leftarrow \left[S_i \ t_i\right]$, and $E_r \leftarrow \left(Q\left(i\right), a_r, b_r\right)$.

$$\begin{aligned} &\max_{\tau,x}\ \ \tau \\ &\text{s.t.,}\ \ S_{Q(i)}x \le t_{Q(i)} + \tau \\ &\qquad a_f^T x = b_f \\ &\qquad S_i x = t_i \\ &\qquad \tau \ge -c, c \in \mathbb{R}_{>0} \end{aligned} \tag{42}$$

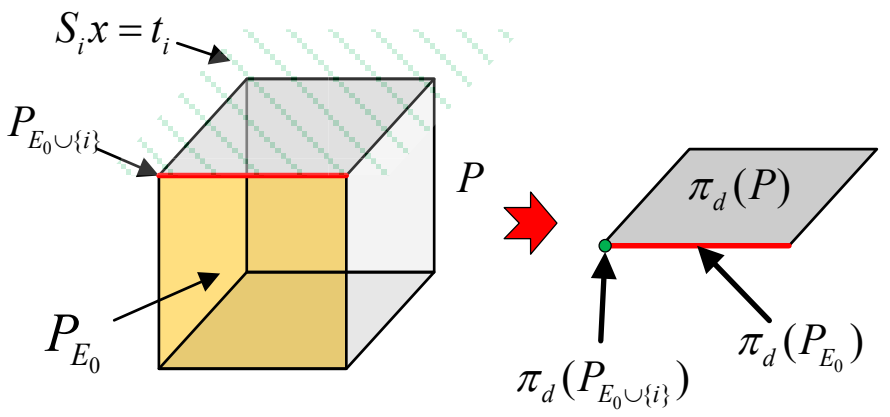

Fig.3. Process of RS

3) *Adjacent Facet Search (AFS)*: The input is two equality sets $E$ and $E_r \supset E$ that define a facet $\pi_d\left(P_E\right)$ and a ridge $\pi_d\left(P_{E_r}\right)$ of the projection $\pi_d\left(P\right)$. The output is an equality set $E^{adj} \subset E_r$ such that $\pi_d\left(P_{E^{adj}}\right)$ is the facet of the projection that is adjacent to the given facet $\pi_d\left(P_E\right)$ and that contains the given ridge $\pi_d\left(P_{E_r}\right)$. The unit normal $a_f$ and $a_r$ are orthogonal:

$$\pi_d\left(P_E\right) = \left\{x \middle| a_f^T x = b_f\right\} \cap \pi_d\left(P\right) \tag{43}$$

$$\pi_d\left(P_{E_r}\right) = \left\{x \middle| a_r^T x = b_r\right\} \cap \pi_d\left(P_E\right) \tag{44}$$

To illustrate the process, we take an example from $\mathbb{R}^d$ to $\mathbb{R}^2$, as shown in Fig. 2. We define an affine transformation $\rho$ that maps $\mathbb{R}^d$ to $\mathbb{R}^2$, in which all points in the polyhedron $\pi_d\left(P\right)$ are mapped to a 2-dimension space with $\alpha$-axis and $\beta$-axis. The facet $\pi_d\left(P_E\right)$ is to mapped to the negative $\alpha$-axis and interior of the polyhedron is strictly above the $\alpha$-axis. The adjacent facet of aff $\pi_d\left(P_{E^{adj}}\right)$ will form a line, whose angle must be less than $180^\circ$ due to convex property.

Because each subset $B$ of $E_r$ is a superset of the given ridge $\pi_d\left(P_B\right) \supseteq \pi_d\left(P_{E_r}\right)$, the affine hull contains the affine hull of the given ridge aff $\pi_d\left(P_B\right) \supseteq$ aff $\pi_d\left(P_{E_r}\right)$. Therefore, the affine hull will either map to the origin, or to a line through the origin. The largest angle $\theta$ with the negative $\alpha$-axis must define the adjacent facet.

The adjacent facet search is transformed to find maximum angle $\theta$ between adjacent facet and the given facet $\pi_d\left(P_E\right)$. The simplest method is to compute the projection $\pi_d\left(\text{aff}\, P_B\right)$ for each $B \subset E_r$ and compute the angle. The largest angle is formulated as a maximize over $\alpha$ whiling fixing positive value $\beta$. The LP can be formulated as follows:

$$\begin{aligned} &\max_{(x,y)\in P}\ \ a_r^T x \\ &\text{s.t.,}\ \ C_{E_r}x + D_{E_r}y \le b_{E_r} \\ &\qquad a_f^T x = b_f(1-\varepsilon) \end{aligned} \tag{45}$$

where $\varepsilon$ is a small positive number used to fix value $\beta$.

If $\left(x^*, y^*\right)$ is an unique optimal solution of LP (45), and $E^{adj} = \left\{i \in E_r \middle| C_i x^* + D_i y^* = b_i\right\}$, then $E^{adj}$ is an equality set of $P$, $\pi_d\left(P_{E^{adj}}\right) \supset \pi_d\left(P_{E_r}\right)$ is a facet of $\pi_d\left(P\right)$, and dim $P_{E^{adj}} = d-1$.

Therefore, affine hulls of adjacent facets can be computed according to **Lemma 4:**

$$\left[a_{adj}^T \ \ b_{adj}\right] \leftarrow \mathrm{N}\left({D_{E^{adj}}}^T\right)^T \left[C_{E^{adj}} \ \ b_{E^{adj}}\right] \tag{46}$$

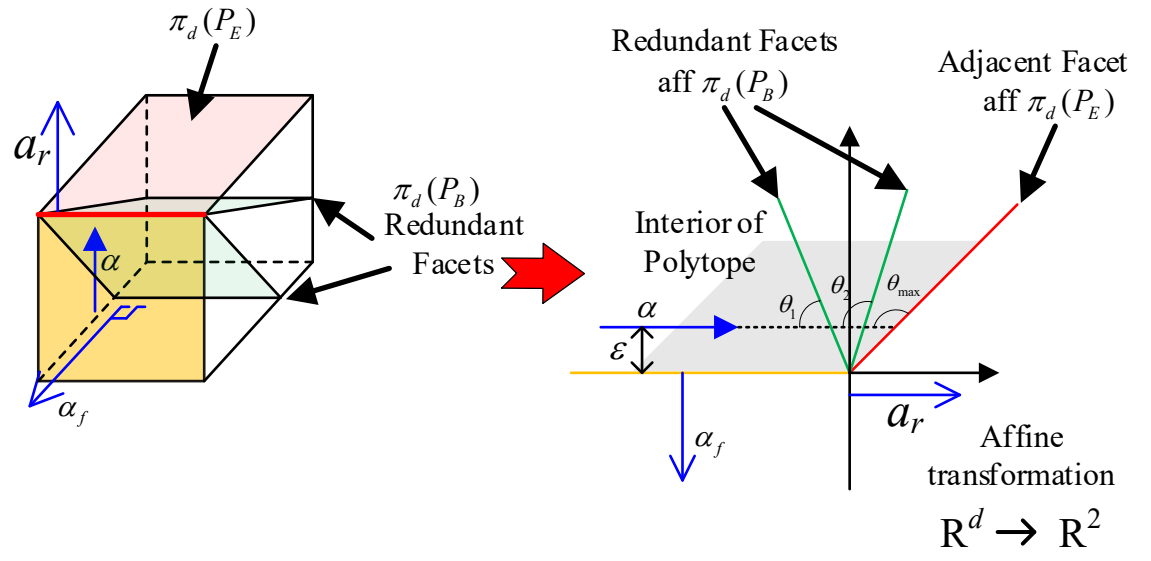

Fig.4. Process of AFS

Fig.5(a) and Fig.5(b) illustrate the FR of a 3-dimension model before and after projection.

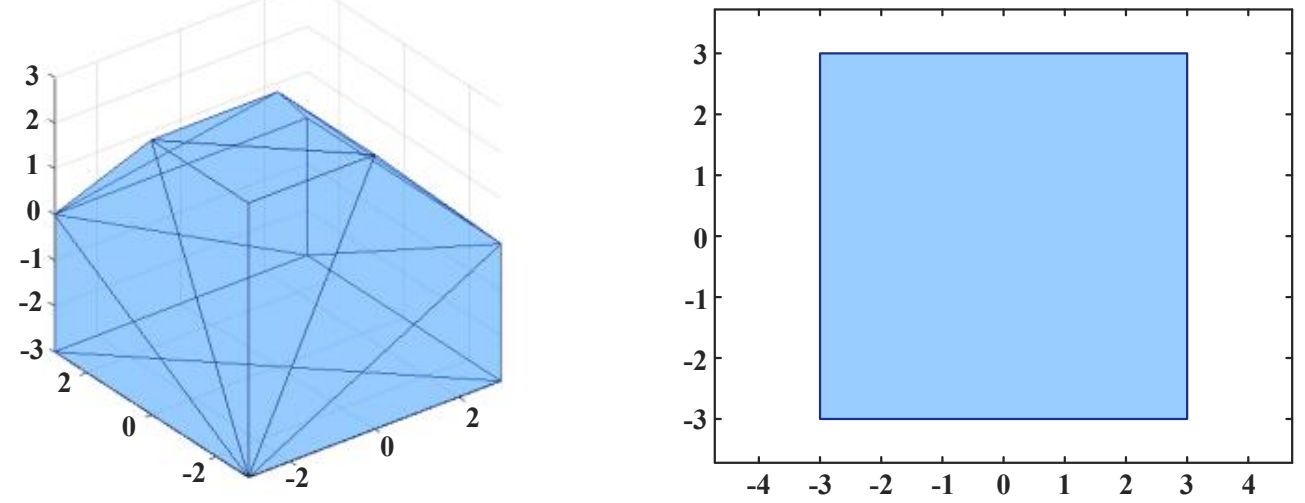

(a) FR before projection (b) FR after projection

Fig.5 Illustration of Projecting a 3-dimension FR to 2-dimension

**Algorithm 1** ESP algorithm

**Input**: Polyhedron $P$

***Initialization ridge-facet list L***

Ridge-facet list $L \leftarrow \emptyset$

$\left(\mathrm{E}_0, a_f, b_f\right) = Initialization\left(P\right)$

$\mathrm{E}_r = RS\left(\mathrm{E}_0, a_f, b_f\right)$

Add element $\left(\left(\mathrm{E}_0, a_f, b_f\right), \left(\mathrm{E}_r, a_r, b_r\right)\right)$ to list $L$

***Initialization matrix G, vector g, and list*** **E.**

$\boldsymbol{G} \leftarrow a_f^T$, $\boldsymbol{g} \leftarrow b_f$, $\mathbf{E} \leftarrow \mathrm{E}_0$

**Search for adjacent facets until the list L is empty**

while $L \neq \emptyset$ do

  Choose an $\left(\left(\mathrm{E}_0, a_f, b_f\right), \left(\mathrm{E}_r, a_r, b_r\right)\right)$ from **L**

  $\left(\mathrm{E}^{\mathrm{adj}}, a_{\mathrm{adj}}, b_{\mathrm{adj}}\right) \leftarrow AFS\left(\left(\mathrm{E}_0, a_f, b_f\right), \left(\mathrm{E}_r, a_r, b_r\right)\right)$

  $\mathrm{E}_r \leftarrow RS\left(\mathrm{E}^{\mathrm{adj}}, a_{\mathrm{adj}}, b_{\mathrm{adj}}\right)$

  if a ridge appears in the list twice

    Remove the ridge

  else

    Add $\left(\left(\mathrm{E}_r, a_r, b_r\right), \left(\mathrm{E}^{\mathrm{adj}}, a_{\mathrm{adj}}, b_{\mathrm{adj}}\right)\right)$ to list **L**

  end if

  $G \leftarrow a_{\mathrm{adj}}^T$, $g \leftarrow b_{\mathrm{adj}}$, $\mathbf{E} \leftarrow \mathrm{E}^{\mathrm{adj}}$

end while

**Output:** $G$, $g$, and **E**

### *C. Accelerated ESP Algorithm*

The original Adjacent Facet Search algorithm is required non-degeneracy situation to ensure a unique optimal solution regarding the LP (45) [23]. Under the non-degeneracy, the Adjacent Facet Search algorithm exhibits well-defined output sensitivity, with its overall computational complexity scaling linearly with the number of projection facets. This uniqueness allows the algorithm to directly identify the equality set of the adjacent facet. However, degeneracy occurs when the optimal solution is not unique, thus requiring supplementary procedures to determine the correct solution. This process requires affine reconstruction and potentially recursive global activeness tests, resulting in significant computational burden. Thus, the affine support of the adjacent facet is first reconstructed from any optimal solution, after which a global activeness test is applied to determine the precise equality set.

Model (47) is employed to determine a unique equality set when dual degeneracy occurs in the original ESP algorithm.

$$\begin{aligned} J(k) &= \min C_k x + D_k y - b_k \\ \text{s.t.} \quad & C_{\mathrm{E}_c^k} x + D_{\mathrm{E}_c^k} y - b_{\mathrm{E}_c^k} \\ & x = x^* \end{aligned} \tag{47}$$

The equality set of $\Omega$ consists of all constraints that hold at every point throughout the polytope. Thus, if $J(k) \neq 0$, constraint $k$ is either excluded from the equality set (when negative) or redundant (when positive); a constraint becomes binding only if $J(k) = 0$. This iterative verification approach significantly increases computational time. Consequently, determining a unique equality set requires a unique solution for $y^*$.

This paper proposed an accelerated ESP algorithm that can directly computes a unique $y^*$ and thereby uniquely fixes the adjacent facet's equality set with a given boundary point $x^*$. The iterative verification in Model (47) is replaced by a convex Quadratic Programming–based Model (48) to eliminate branching under degeneracy.

$$\begin{aligned} & \min_y \frac{1}{2} y^{\mathrm{T}} \Lambda y - d^{\mathrm{T}} y \\ & s.t. \;\; Cx^* + Dy \le b \\ & \quad C_{E_r} x^* + D_{E_r} y = b_{E_r}(1-\varepsilon) \end{aligned} \tag{48}$$

where $\Lambda \succ 0$, and $\frac{1}{2} y^{\mathrm{T}} \Lambda y$ is $\ell 2$ regularization term. Model (48) is a quadratic programming model when $d = 0$, which is equivalent to the Euclidean projection of the origin onto the feasible slice:

$$Y(x^*) = \{ y \in \mathbb{R}^k \mid Cx^* + Dy \le b \} \tag{49}$$

Equation (49) is a closed convex polyhedron. A unique global solution $y^*$ to the model exists when the reduced-Hessian matrix $y^T \Lambda y$ is positive definite. Computational efficiency is enhanced by avoiding iterative process within model (47) through a single execution of model (48).

The uniqueness of the solution for Model (48) is proofed as follows.

*Proof:*

The adjacent-facet search model can be written as the following quadratic program:

$$\begin{aligned} & \min q(x_L) = \frac{1}{2} x_L^{\mathrm{T}} G_L x_L + x_L^{\mathrm{T}} c_L \\ & \text{s.}t. \quad \mathrm{A}_L^{\mathrm{T}} x_L \le b_L \end{aligned} \tag{50}$$

If problem (50) has a feasible solution, based on Karush-Kuhn-Tucker(KKT) conditions, we obtain:

$$\begin{bmatrix} G_L & A_L^{\mathrm{T}} \\ A_L^{\mathrm{T}} & 0 \end{bmatrix} \begin{bmatrix} x_L \\ \lambda_L \end{bmatrix} = \begin{bmatrix} -c_L \\ -b_L \end{bmatrix} \tag{51}$$

where, $\lambda_L$ denotes the vector of Lagrange multipliers. here exist vectors $w_L$ and $v_L$ such that:

$$\begin{bmatrix} G_L & A_L^{\mathrm{T}} \\ A_L^{\mathrm{T}} & 0 \end{bmatrix} \begin{bmatrix} w_L \\ v_L \end{bmatrix} = 0 \tag{52}$$

since $A_L w_L = 0$, therefore:

$$0=\begin{bmatrix} w_L^{\mathrm{T}} & v_L^{\mathrm{T}} \end{bmatrix}\begin{bmatrix} G_L & A_L^{\mathrm{T}} \\ A_L & 0 \end{bmatrix}\begin{bmatrix} w_L \\ v_L \end{bmatrix}=w_L^{\mathrm{T}} G_L w_L \tag{53}$$

Because $w_L$ lies in the null space of $A_L$, it can be written as $w_L = Zu_L$, $u_L \in R^{n-m}$. Hence:

$$0 = w_L^{\mathrm{T}} G_L w_L = u_L^{\mathrm{T}} Z^{\mathrm{T}} G_L Z u_L \tag{54}$$

Since $Z^{\mathrm{T}} G_L Z$ is positive definite, it follows that $u_L = 0$ and thus $w_L = 0$. It follows formular (52) that $A_L^{\mathrm{T}} v_L = 0$. because $A_L$ has full row rank, this implies $v_L = 0$. Therefore, the only vectors satisfying (52) are $w_L = 0$ and $v_L = 0$, which means the coefficient matrix is nonsingular. Consequently, there exists a unique solution $(x_L^*, \lambda_L^*)$ to the first-order KKT conditions in (51), and $x_L^*$ is the unique global minimizer of the quadratic programming problem.

### *D. Illustrative example of ESP algorithm*

The following example illustrates the process of a simple 3-dimension FR is defined as:

$$P = \{(x_1, x_2, y) \mid 0 \le x_1 \le 1, 0 \le x_2 \le 1, 0 \le y \le 1\} \tag{55}$$

where $P$ is a 3-dimension polytope. $x_1$, $x_2$ represent the variables that will remain, $y$ represents the variable that will be eliminated. Formula (55) could be written as the formulation below:

$$P = \{(x, y) \mid Cx + Dy \le b\}, x = (x_1, x_2) \in \mathbb{R}^2 \tag{56}$$

Step 1: An initial facet could be found by Initialization algorithm. Therefore, the 2-dimension affine hull projection is:

$$\mathrm{aff}\pi_2(P_{E_0}) = \{x \mid a_f^{\mathrm{T}} x = b_f\} \tag{57}$$

where $a_f = [1, 0]^{\mathrm{T}}, b_f = 1$. Then, a 2-dimension projection of $P_{E_0}$ is:

$$\pi_2(P_{E_0}) = \{x \mid x_1 = 1, 0 \le x_2 \le 1\}, E_0 = \{1\} \tag{58}$$

Step 2: RS is employed to find the ridges of $\pi_2(P_{E_0})$. The equality set of ridges could be constructed by model (42). In this example, they are $E_{r_1} = \{1, 3\}$ and $E_{r_2} = \{1, 4\}$.

Step 3: AFS is used to find the adjacent facets of $\pi_2(P_{E_0})$. The equality sets of the adjacent facets could be calculated via model (45). Each ridge corresponds to an adjacent face. The equality sets of adjacent faces are $E_1 = \{3\}$ and $E_2 = \{4\}$.

After Step 3 is completed, all the adjacent facets of $\pi_2(P_{E_0})$ are found. Step 2 and Step 3 should be repeated until all newly found ridges are repeated with the existing ridges. The process of Step 1, Step 2 and Step 3 is demonstrated in Fig.6.

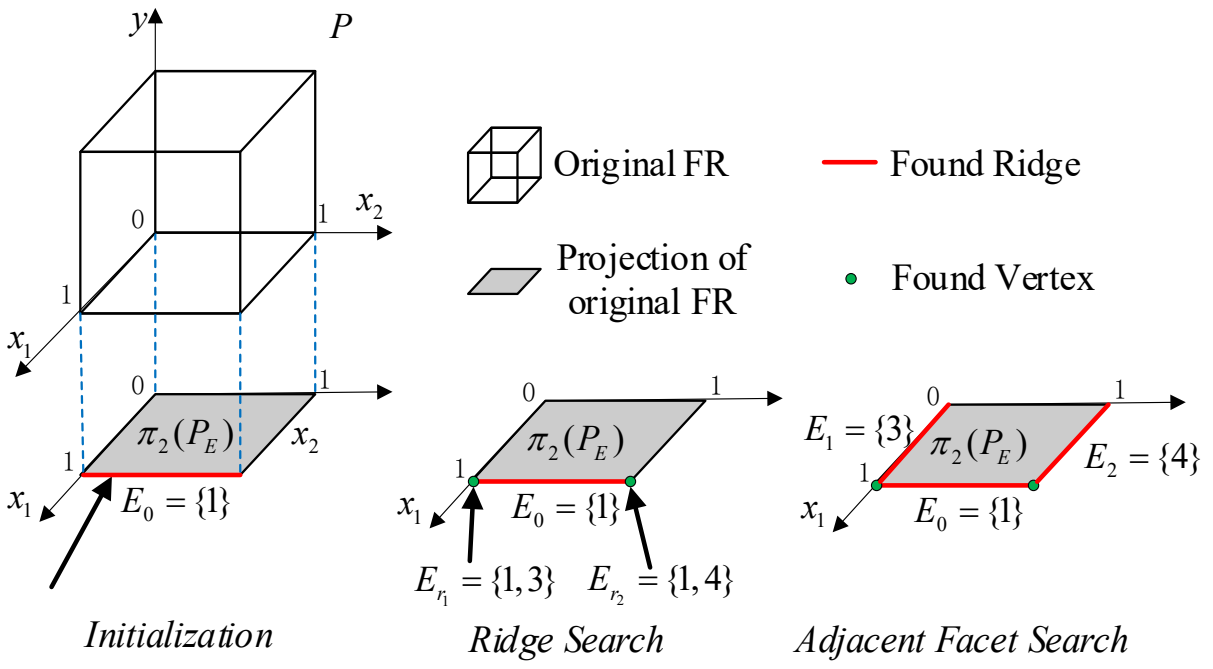

Fig. 6. An example of a projection process

## IV. Case Studies

In this section, three cases, including a test dataset of different polyhedrons, IEEE 33-bus system, and T118D10 TSO-DSO system, are used to verify the effectiveness of the proposed methods. YALMIP and GUROBI 11.0.1 are used to model and solve the proposed model.

### *A. Analysis of computational efficiency of accelerated ESP algorithm*

The computational efficiency of accelerated ESP algorithm is validated based on a test dataset constructed by a series of polyhedrons, including 10-dimensions, 13-dimensions, 15-dimensions, 17-dimensions, 20-dimensions, and 25-dimensions polyhedrons. Further, three 10-dimensional polyhedrons with varying dual degeneracy frequencies during projection are included to evaluate the ESP algorithm and accelerated ESP algorithm under degenerate conditions. All polyhedrons are defined by constraint sets consisting of 70% box constraints and 30% linear constraints. The target of dimension after projection is set to 7.

The process of VE algorithm is demonstrated in Fig.7. First, in Fig.7(a), vertex A and B are identified after two vertex enumeration processes, which are shown in Fig.7(a). Second, in Fig.7(b), the normal vector of the straight line between A and B are taken as the search direction, then vertex C is found. Finally, in Fig.7(c), the normal vector of the straight line between A and C are taken as the search direction, then the last vertex D is found.

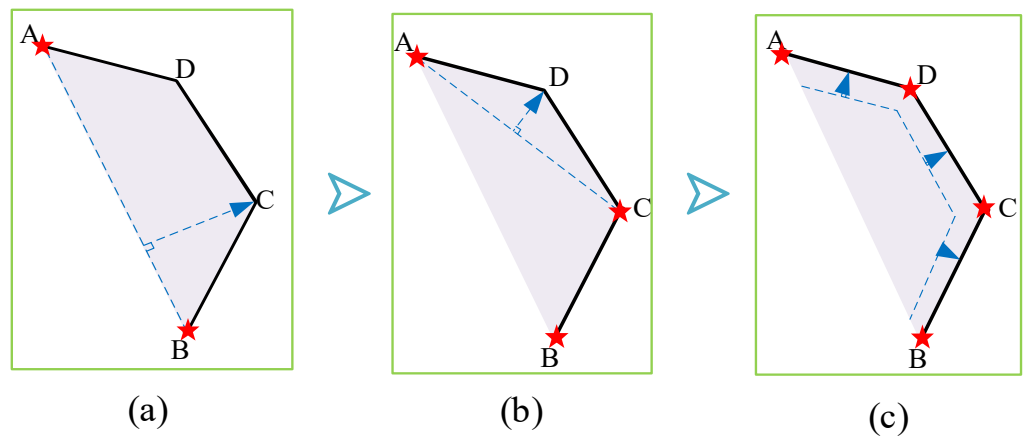

Fig. 7. Process of VE algorithm

For accelerated ESP algorithm, it shows equal computational efficiency to ESP algorithm when dual degeneracy does not occur. Meanwhile, both ESP algorithm and accelerated ESP algorithm are more efficient than VE algorithm. Furthermore, the curse of dimensionality faced by VE algorithm in 25-dimension non-degeneracy case is avoided in both ESP algorithm and accelerated ESP algorithm. This is because the computational burden of VE algorithm is highly sensitive to the number of vertices. Each additional dimension or active constraint could introduce large number of vertices. ESP algorithm constructs the FR by facet-to-facet searching, which makes it less sensitive to the increase in vertices. The computational efficiency of ESP algorithm, accelerated ESP algorithm and VE algorithm on non-degeneracy scenarios are shown in Fig. 8

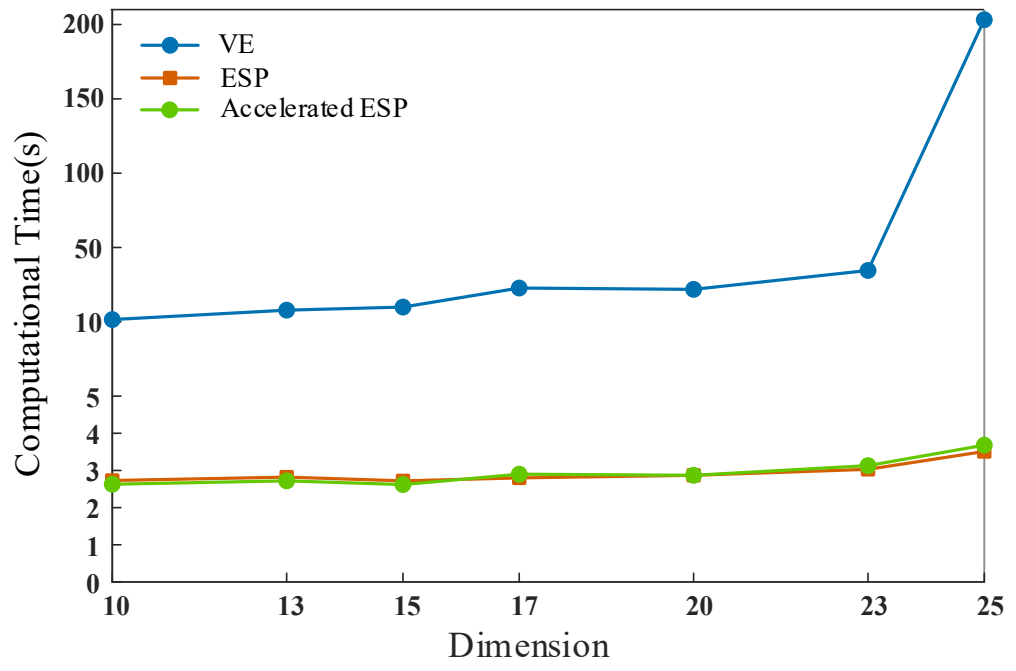

Fig. 8. Computational efficiency of VE algorithm, ESP algorithm and accelerated ESP algorithm in different non-degeneracy cases

On dual degeneracy occasions, the accelerated ESP algorithm achieves a 4.2% to 61.0% efficiency improvement to ESP algorithm. This efficiency advantage increased as the number of dual degeneracies increased. This is because the occurrence of dual degeneracy could lead to recursion and constrain enumeration, thereby enlarging the computation burden. The computational efficiency of ESP algorithm and accelerated ESP algorithm on dual-degeneracy scenarios are shown in Fig. 9

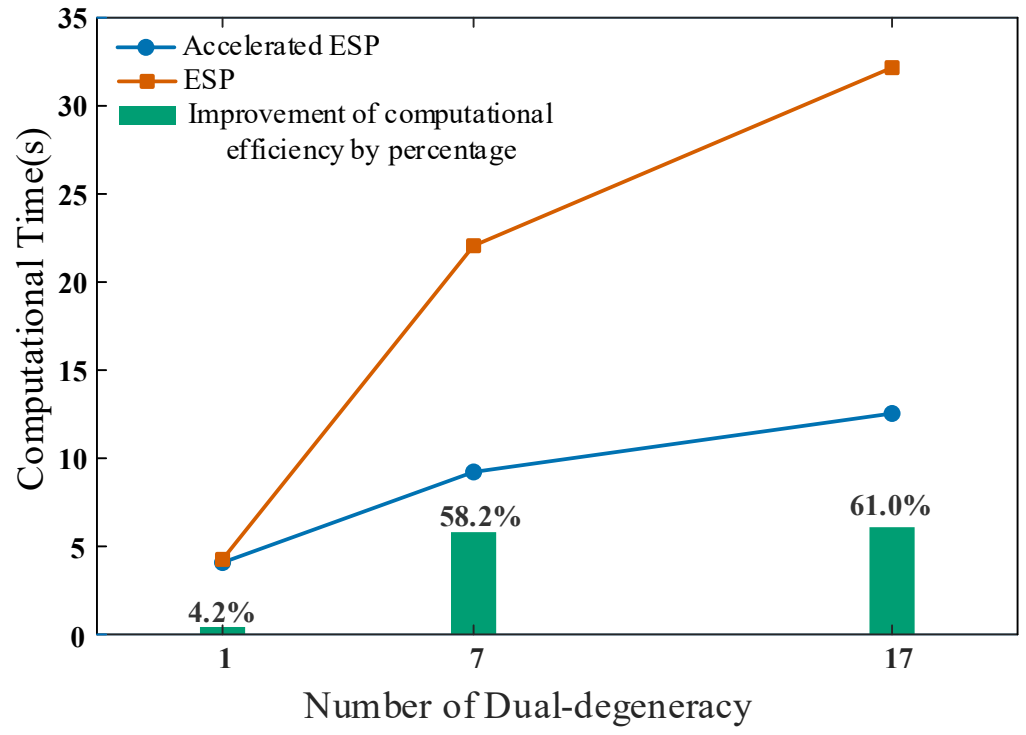

Fig. 9. Computational efficiency of ESP and accelerated ESP in different 10-dimenison dual degeneracy cases

## B. IEEE 33-Bus System

The IEEE 33-bus system includes photovoltaic (PV), wind turbine (WT), and gas turbine (GT) units. WT locates on bus 21, 29, and 32. PV locates on bus 10, 15, and 18. GT locates on bus 8, 14, 24, and 28. The installed capacity of WT, PV, and GT are 10MW, 11MW, and 20MW, respectively. Generation profiles of PV and WT units are from Ref.[24]. The simulation time instance is one day with 24 hours. The topology of the IEEE 33-bus system is demonstrated in Fig. 10. The output curve of renewable energy and the load curve is demonstrated in Fig. 11.

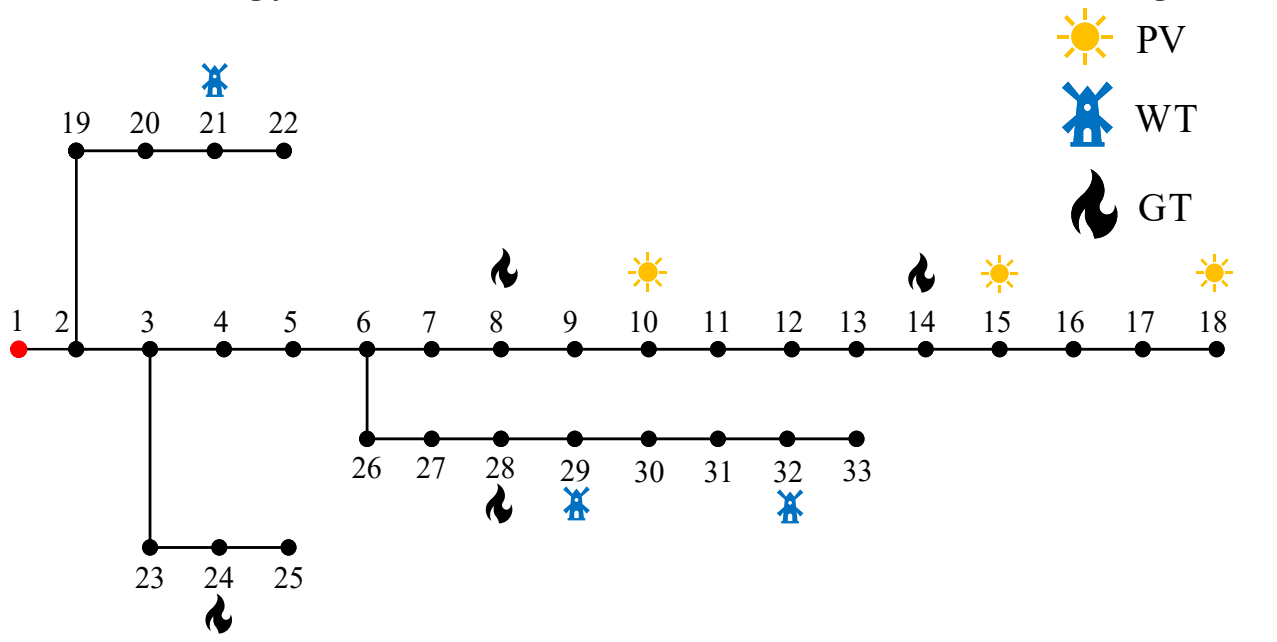

Fig. 10. Topology of IEEE 33-bus system

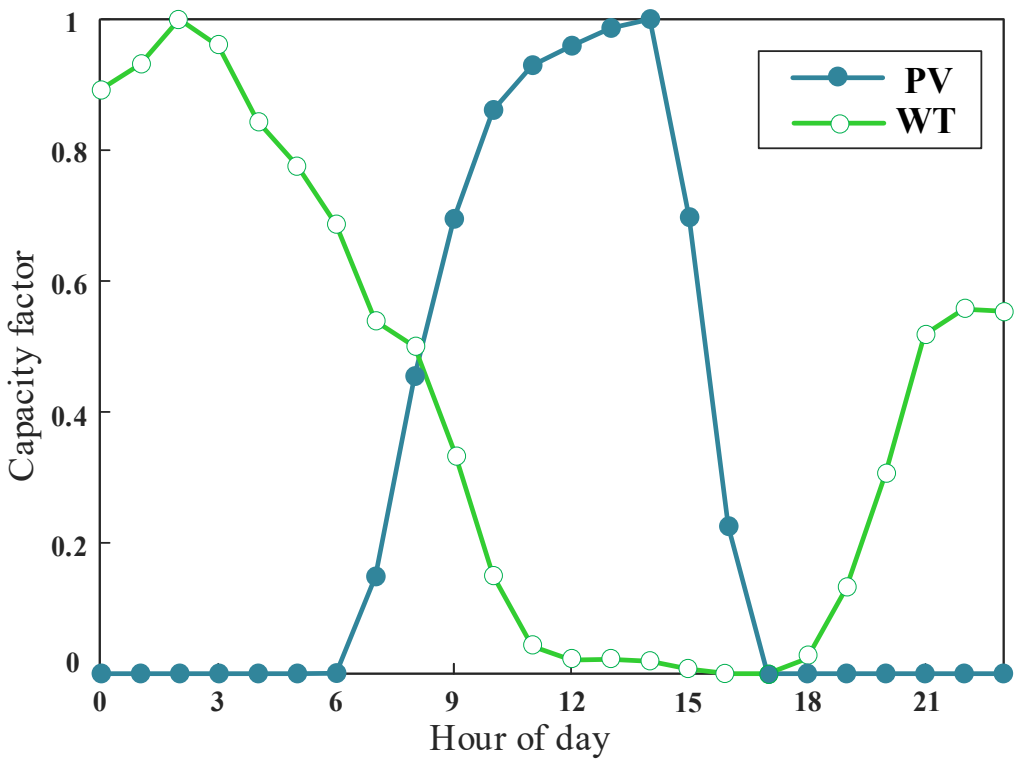

Fig. 11. The output curve of renewable energy and the load curve

Three EP algorithms are compared in terms of accuracy and computational efficiency.

S1: VE algorithm [12], [13].

S2: original ESP algorithm.

S3: accelerated ESP algorithm.

For both the three-dimensional (3-D) FR and six-dimensional (6-D) FRs, the projection computation times of S2 and S3 are shorter than that of S1. Specifically, for the 3-D FR, the projection computation times of S2 and S3 are identical, while that of S1 is 60.02% longer than both S2 and S3. For the 6-D FR, the projection computation time of S1 is 67.13% longer than that of S2 and 205.15% longer than that of S3. This is because the VE algorithm requires traversing not only all vertices but also all facets during the verification phase. The verification procedure involves constructing facets between adjacent vertices and performing directional searches along their respective normal vectors. The absence of new vertices is confirmed when all searched points fall within the preset distance thresholds. Consequently, the complexity of the VE algorithm scales linearly with the number of vertices and facets.

Table II. Computation time in different dimensions

| Algorithms | 3-D (s) | 6-D(s) |
|---|---|---|
| S1 | 15.3 | 263.4 |
| S2 | 9.6 | 157.6 |
| S3 | 9.6 | 86.3 |

For comparison between S2 and S3, the computational efficiency of both ESP algorithms was identical when the 3-D subspace was projected. However, when projecting the 6-D subspace, the projection computation time of original ESP algorithm is 82.59% higher than the accelerated ESP algorithm. This can be attributed to the increased prevalence of degeneracy phenomena in high-dimensional models. If there exists degeneracy, it is required for hyperplane to be traversed using model (47) in the original ESP algorithm. In contrast, a unique global solution is directly obtained through model (48) in the proposed ESP algorithm, resulting in enhanced computational efficiency.

## C. T118D10 TSO-DSO System

The T118D10 TSO-DSO system consists of an IEEE 118-bus system and ten IEEE 69-bus systems. WTs are connected to buses 10, 15, 49, 87, and 89 of DSO. The DSOs are interconnected with buses 3, 10, 11, 27, 32, 44, 59, 76, 78, and 101 of TSO. Generation profiles of PV and WT units in are from Ref. [24], and the technical data such as the capacity and climbing rate of coal-fired units, and the parameters of lines could be found in [25]. To verify the impact of incorporating distribution network equivalent models on COD, two comparative schemes are established accordingly.

M1: TSO-DSO coordination dispatch.

M2: TSO and DSO dispatch independently.

Table III demonstrates the computational result of the operation cost. For T118D10 system, M1 delivers better system-wide economics and higher renewable utilization than M2. This advantage stems from power complementarity between the TSO and DSO.

Specifically, under M2, the transmission and distribution networks are optimized independently, so the DSO's daily operating cost is 47.59% lower than under M1. This is because it need not accommodate exporting surplus renewable generation and can focus on internal resource allocation. However, this local optimum forces the transmission grid to ramp high-cost thermal units to cover power deficits at certain times, which must ramp high-cost thermal units to cover deficits at certain times. Thus, the total cost of M2 is 5.29% higher than M1. Under M1, the operation cost of DSO rises by 90.81% due to increased gas-turbine output, but EP-based TSO-DSO coordination allows more flexible dispatch and dynamic tie line control while considering the renewable generator within DSO renewables. Due to high renewable energy utilization rate, transmission penalty cost reduced 5.03% and distribution curtailment by 40.44%, yielding a 14.2% reduction in total system cost. Consequently, the overall economic efficiency of TSO–DSO coordination operation is improved.

Table III. Result of dispatch models (unit: $10^3$ USD)

| Scenarios | TSO Cost | RE curt.(T) | DSO Cost | RE curt.(D) | Total Cost |
|---|---|---|---|---|---|
| M1 | 1816.2 | 192.5 | 76.7 | 508.0 | 2593.4 |
| M2 | 1912.3 | 202.8 | 40.2 | 867.4 | 3022.6 |

## V. Conclusion

In this paper, an accelerated ESP algorithm-based non-iterative TSO-DSO coordination approach is proposed to overcome the bottleneck in projection construction of high-dimension FR. In detail, a reliable method for efficiently constructing the low-dimension FR of DSO dispatch model is developed to avoid the occurrence of unsolvable cases using existing solvers. A $\ell 2$ regularization-based method is employed to avoid the recursive process causes by dual degeneracy. The case study shows that for the 6-D model, the computation time of the proposed approach is reduced by more than 45.24% compared with benchmark methods. The developed FR characterization method can be extended to the DERs aggregation problem in other COD and electricity market scenarios.